\documentclass[graybox]{svmult}

\usepackage{type1cm}        % activate if the above 3 fonts are
\usepackage{makeidx}         % allows index generation
\usepackage{graphicx}        % standard LaTeX graphics tool
\usepackage{multicol}        % used for the two-column index
\usepackage[bottom]{footmisc}% places footnotes at page bottom

\usepackage{cite}         % collapses citations

\usepackage{newtxtext}       % 
\usepackage[varvw]{newtxmath}       % selects Times Roman as basic font

\makeindex             % used for the subject index
\newcommand{\Trento}{T$_{\rm{R}}$ENTo}

\begin{document}

\title*{
Hydrodynamic Description of the\\ Quark-Gluon Plasma}

\author{Ulrich Heinz and Bj\"orn Schenke}

\institute{Ulrich Heinz \at  Physics Department, The Ohio State University, Columbus, OH 43210, USA;\\  \email{heinz.9@osu.edu}
\and Bj\"orn Schenke  \at  Physics Department, Brookhaven National Laboratory, Upton, NY 11973, USA;\\ \email{bschenke@bnl.gov}}

\maketitle

\abstract{
We review the history and success of applying relativistic hydrodynamics to high-energy heavy-ion collisions. We emphasize the important role hydrodynamics has played in the discovery of the quark-gluon plasma and its quantitative exploration.
}

\section{Introduction and Historical Overview}
One of the most impactful discoveries made by studying high energy heavy-ion collisions has been that of the nearly perfect (or inviscid) fluidity of the matter created in such collisions. Based on a series of observables, designed to determine the nature of the created matter \cite{Harris:1996zx}, it was established that the nearly perfect fluid is a quark-gluon plasma (QGP) \cite{Muller:2012zq, Harris:2024aov}. This QGP does not have the typical hadronic degrees of freedom, but instead is a system of liberated quarks and gluons which, as its fluid characteristic implies, are strongly interacting.

How did we arrive at this conclusion and what role did relativistic hydrodynamics play? Let us back up by 50 years to the 1970s when heavy-ion collisions were proposed to study the question of `vacuum' and how to excite it, and as a tool for achieving abnormal conditions of nuclear matter. They allowed to investigate `bulk' phenomena and the condensed matter properties of strongly interacting matter by distributing high energy and nucleon density over a relatively large volume. An important event where these questions were discussed was the Bear Mountain, NY, workshop held in November/December of 1974 \cite{osti_4061527, Baym:2001in}.

Early experimental setups, like those at the BEVALAC, AGS, and CERN’s SPS (see e.g. \cite{Stoecker:1986ci} for an early review), laid the groundwork for larger-scale facilities such as the Relativistic Heavy Ion Collider (RHIC) at Brookhaven National Laboratory (BNL) and the Large Hadron Collider (LHC) at CERN, which have enabled the creation and detailed study of the QGP \cite{Shuryak:2014zxa, Busza:2018rrf, Harris:2024aov}. 

The theoretical foundation for these studies builds on relativistic hydrodynamics, a framework introduced for high-energy collisions by Landau in 1953 \cite{Landau:1953gs}. Following early works by Cattaneo \cite{cattaneo}, relativistic equations of motion for viscous fluids were derived in the 1960s and 70s by M\"uller, and Israel and Stewart \cite{Muller:1967zza, ISRAEL1976310, Israel:1976, Israel:1979wp}. These included viscous corrections derived from thermodynamic principles and avoided problems with causality (which also result in numerical instability) that the relativistic Navier-Stokes equations encounter \cite{Romatschke:2009im}.

In the 1980s, Bjorken provided a simple framework for analytically modeling heavy-ion collisions \cite{Bjorken:1982qr} under the assumption of boost invariance, or the existence of a central plateau in the particle production as a function of rapidity. This model was borne out as a useful approximation by SPS collider data. The approximate independence of rapidity near the midrapidity region is often used to this day to simplify numerical calculations by reducing the spatial dimensions to the two transverse ones.

First successes of the fluid dynamical approach in providing a qualitative understanding of key features of the rapidly growing body of experimental data at the BEVALAC, AGS and SPS spawned a flurry of theoretical activities in developing hydrodynamic frameworks for heavy-ion collisions \cite{Scheid:1974zz, Baumgardt:1975qv, Stoecker:1986ci, Clare:1986qj}, hydrodynamically inspired semi-analytic models \cite{Siemens:1978pb, Schnedermann:1992hp}, and numerical solutions invoking approximate symmetries to reduce computational complexity (e.g. \cite{Baym:1983amj, VonGersdorff:1986tqh, Ornik:1989jp, Staubo:1989wa}). 

Acceptance of the hydrodynamic approach was initially slow due to the unresolved question how the small and short-lived fireballs could thermalize fast enough for fluid dynamics to become applicable. Motivated by such questions, the 1980s and 90s also brought investigations of QCD-based kinetic theory \cite{Heinz:1983nx, Elze:1989un} as well as the first quantum field theoretical computations of transport coefficients such as the shear and bulk viscosities \cite{Hosoya:1983xm, Baym:1990uj, Heiselberg:1994vy, Jeon:1994if, Jeon:1995zm}, which in principle are important inputs into the hydrodynamic equations. However, as such calculations are feasible mainly in the extreme limits of weak \cite{Arnold:2000dr} or strong \cite{Policastro:2001yc} coupling (which are not good approximations under real-life conditions encountered in heavy-ion collisions), the most efficient approach to determining transport coefficients has been to constrain them by comparison of hydrodynamic model results with experimental data. We will discuss this approach as well as modern developments in computing transport coefficients within different theories below. 

The 1990s saw much development in hydrodynamically motivated models such as the blast wave model \cite{Schnedermann:1992hp, Schnedermann:1993ws}, algorithm developments for solving the hydrodynamic equations numerically \cite{Rischke:1995ir, Schlei:1995hv, Sollfrank:1996hd, Teaney:1999gr, Kolb:1999it}, as well as first hybrid models that combined hydrodynamics with a microscopic description of the more dilute late hadronic stage of the collision \cite{Dumitru:1999sf, Bass:1999tu, Bass:2000ib, Hirano:2005wx}. An important change of paradigm characterized these developments: whereas originally the hydrodynamic approach was applied to the entire collision process, from the initial approach of the two colliding nuclei all the way to the final freeze-out of the hot medium produced in the collision, contemporary models split the evolution history into several stages that are described with different theoretical tools: 
\begin{itemize}
    \item 
    an initial far-off-equilibrium energy deposition and ``hydrodynamization'' stage, that must be described microscopically, followed by 
    \item 
    the collective expansion of a hot and dense medium made of quark-gluon plasma or, at lower temperatures, of hadrons, that is strongly coupled and sufficiently close to local thermal equilibrium for dissipative fluid dynamics to be applicable, and finally 
    \item 
    a late dilute and much more weakly interacting hadronic rescattering and freeze-out stage in which the hydrodynamic fields are converted into individual hadronic resonances (``particlization''), which are further propagated until freeze-out via a hadronic cascade simulating a set of coupled Boltzmann equations for the hadron distribution functions.
\end{itemize} 

Following earlier work on anisotropies in the flow patterns of emitted particles discovered in the 1970s and 1980s in experiments performed at the BEVALAC (``side-splash" and ``squeeze-out'' \cite{Stoecker:1986ci}), the measurement of azimuthal anisotropy in produced particle momentum distributions was proposed in 1992 as a signature of transverse collective flow \cite{Ollitrault:1992bk}. From then on, a lot of focus was on elliptic flow \cite{Voloshin:1994mz, Poskanzer:1998yz}, which in the hydrodynamic picture was expected to be generated by the system's response to the initial anisotropic, mainly elliptic, shape of the produced matter. Two overlapping highly energetic nuclei, assumed to be flat pancakes in the longitudinal direction and round disks in the transverse direction, will only have a round interaction region for zero impact parameter. In general, they have an almond shaped interaction region whose ellipticity increases with impact parameter. This leads to a characteristic prediction for the centrality dependence of elliptic flow in heavy-ion collisions.

\begin{figure}[htbp]
\begin{center}
\includegraphics[width=10cm]{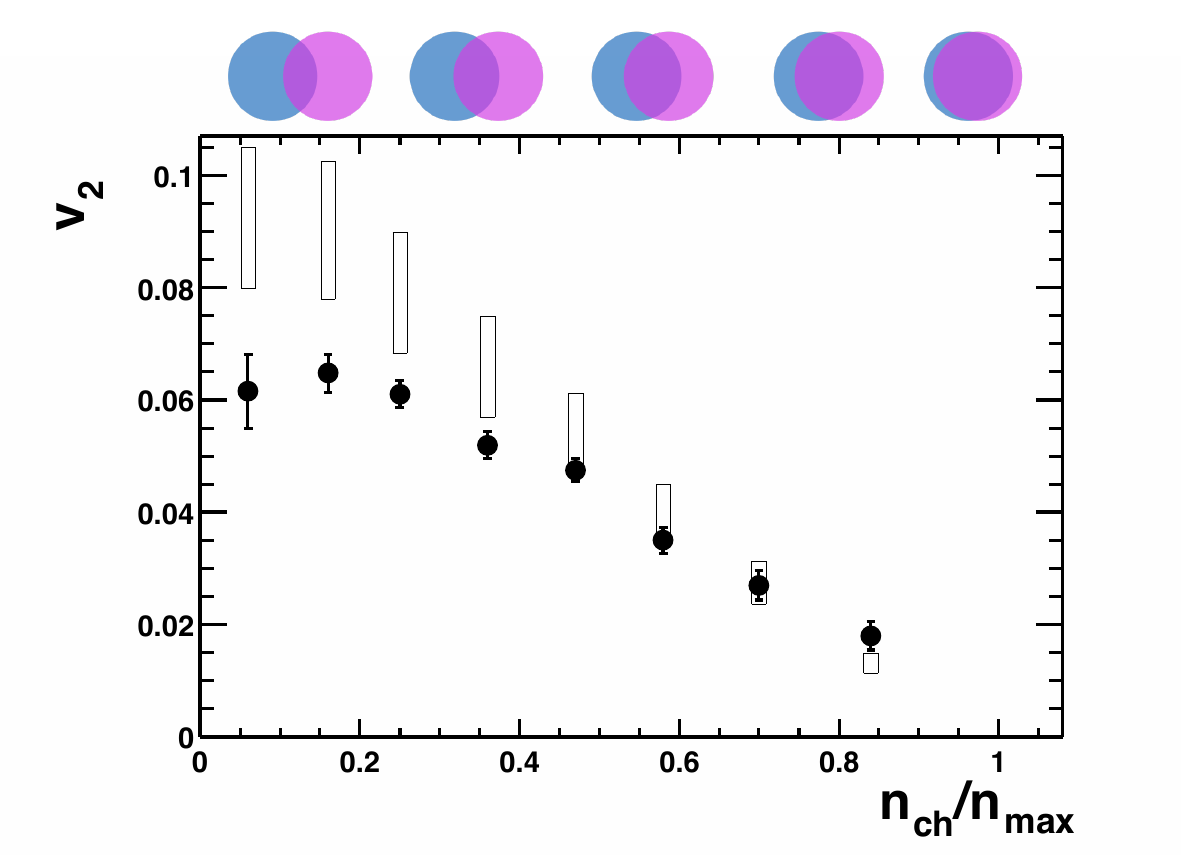}
\end{center}
\caption{The elliptic flow coefficient $v_2$ as a function of centrality, represented by the scaled charged particle multiplicity $n_{\rm ch}/n_{\rm max}$ , as measured by the STAR Collaboration. The open rectangles show expectations for $v_2$ in the ideal hydrodynamic limit. The sketch on top illustrates the overlap of the colliding nuclei in the transverse plane, with decreasing impact parameter as $n_{\rm ch}$ increases. Figure adapted from \cite{STAR:2000ekf}.}
\label{fig:v2STAR}     
\end{figure}

Evidence for directed and elliptic anisotropic flow was already seen in the late 1990s in 160 GeV/nucleon Pb+Pb collisions studied at the CERN SPS \cite{Heinz:2000bk}, but a quantitative understanding of the observations remained elusive. However, when RHIC operation began in 2000, some of the very first results showed unambiguously the expected centrality dependence of the elliptic flow observable $v_2$ in Au+Au collisions \cite{STAR:2000ekf}, with absolute values just slightly lower than the expectations from (ideal) hydrodynamics, as shown in Fig.\,\ref{fig:v2STAR}. After this measurement it became quickly clear that the matter created at RHIC evolved like an almost ideal fluid and thus had to be strongly coupled \cite{Heinz:2001xi, Gyulassy:2004zy, Shuryak:2004cy}.

By 2005 all RHIC experiments had collected enough evidence of the formation of a QGP phase in Au+Au collisions \cite{BRAHMS:2004adc, PHOBOS:2004zne, STAR:2005gfr, PHENIX:2004vcz}, including the elliptic flow, whose magnitude, centrality, and transverse momentum dependence was well described by hydrodynamic models, that Brookhaven National Laboratory published a press release announcing the discovery of the perfect liquid QGP.\footnote{An assessment of the insights gained from the heavy-ion program at the CERN SPS during the 1980s and 1990s \cite{Heinz:2000bk} concluded that compelling evidence for the creation of ``a new form of matter" had been found but stopped short of claiming unambiguous discovery of the quark-gluon plasma, nor did it comment on its perfectly liquid collective dynamical properties. The latter became only obvious after theory had progressed to a quantitative understanding of the bulk of the very comprehensive and precise experimental data collected at RHIC.}

In the meantime, first 3+1D ideal hydrodynamic calculations emerged \cite{Hirano:2001eu, Hirano:2002ds}, and the first phenomenological estimates of the effect of shear viscosity on flow observables were performed \cite{Teaney:2003kp}. First full numerical simulations in 2+1 dimensions provided results on elliptic flow \cite{Romatschke:2007mq, Chaudhuri:2007qp, Song:2007fn, Dusling:2007gi} in 2007, and comparison to RHIC data demonstrated that the shear viscosity to entropy density ratio, $\eta/s$, could not be much larger than a few times the lower bound of $1/(4\pi)$, derived at strong coupling in 2001 \cite{Policastro:2001yc}, or the elliptic flow would be underestimated too much. 

An effective field theory description of fluid dynamics was developed in the late 2000s \cite{Baier:2007ix, Bhattacharyya:2007vjd}, and since then many studies of the hydrodynamic limits of many theories have been performed. This resulted in insights into the applicability of hydrodynamics, particularly in small systems, and the role of hydrodynamic attractors and the dominance of hydrodynamic over transient modes \cite{Florkowski:2017olj}.

Fluctuations of the initial geometry and their impact on event-by-event fluctuations in the collective flow patterns of heavy-ion collisions were first seriously considered in the early 2000s \cite{Drescher:2000ec, Miller:2003kd}, with numerical implementations in hydrodynamic codes following later \cite{Takahashi:2009na}. In 2010, it was demonstrated that such initial state fluctuations are responsible for the generation of odd harmonics, most prominently triangular flow, $v_3$ \cite{Alver:2010gr}. The reason for this is that fluctuations, for example rooted in the quantum fluctuations of nucleon positions in the incoming projectile and target, break the symmetry of the initial overlap region, leading to finite eccentricities of all harmonic orders that fluctuate from collision to collision. 

The year 2010 brought many more important developments. First, the LHC began operation, providing measurements of anisotropic flow in high-multiplicity proton-proton collisions \cite{CMS:2010ifv}. This was an entirely unexpected discovery and triggered research into collectivity in small systems that is still going strong today. Hydrodynamic simulations including viscosity and event-by-event initial state fluctuations advanced to 3+1 dimensions and could now provide realistic computations of all flow harmonics as functions of both particle transverse momentum and rapidity \cite{Schenke:2010rr}. 
Furthermore, a 2D analytic model ---an extension of Bjorken's model--- was introduced by Gubser \cite{Gubser:2010ze}, allowing to also consider transverse flow. 

Finally, 2010 brought the onset of anisotropic hydrodynamics (aHydro) \cite{Florkowski:2010cf, Martinez:2010sc}, designed to account non-perturbatively for the large anisotropy between the longitudinal and transverse pressure caused by the initially very large difference between the longitudinal and transverse expansion rates of the created medium. aHydro was the first attempt to extend the range of validity of relativistic fluid dynamics into the domain of large viscous pressures, by using a moment expansion of the Boltzmann equation based on a perturbative expansion of the distribution function around an anisotropic (rather than isotropic local equilibrium) momentum distribution. The form of the latter was based on an approximate solution of the Boltzmann equation for Bjorken flow (which dominates the flow pattern in high-energy collisions at early times). 

The LHC delivered many exciting results in Pb+Pb collisions \cite{ALICE:2011ab,CMS:2011cqy,ATLAS:2012at,Muller:2012zq,ALICE:2018rtz, ATLAS:2018ezv, ALICE:2022wpn}, and in 2012 also provided flow studies in the intermediate and asymmetric p+Pb collisions \cite{CMS:2012qk, ALICE:2012eyl, ATLAS:2012cix, CMS:2013jlh}, finding significant azimuthal anisotropies, as it had in p+p collisions. Theoretical research on many fronts tried to explain those anisotropies. Major candidates were initial state correlations from e.g.~the color glass condensate, but also final state effects, either described by kinetic theory or hydrodynamics \cite{Dusling:2015gta}. Fluid dynamic calculations were applied to a wide range of systems in \cite{Weller:2017tsr}, finding reasonable agreement with observables in Pb+Pb, p+Pb, and even p+p collisions at LHC energies. 

The experimental program at RHIC set out to see whether final state effects are the main origin of the observed anisotropies in small systems by performing a system scan. Here, p+Au, d+Au, and $^3$He+Au collisions were chosen because one expected different initial geometries, with e.g.~triangular structures enhanced for $^3$He+Au collisions, where one has three nucleons in one projectile. PHENIX measurements confirmed \cite{PHENIX:2018lia} the expectation from geometry, although recent studies indicate that subnucleon fluctuations and differences in the longitudinal structure alter the interpretation of the results somewhat \cite{STAR:2022pfn,Zhao:2022ugy}. Nevertheless, current consensus is that strong final state interactions are required to describe the data not only in heavy-ion but also in smaller collision systems. For more detail on collectivity in small systems see \cite{Nagle:2018nvi, Schenke:2021mxx, Noronha:2024dtq, Grosse-Oetringhaus:2024bwr}.

The seemingly unreasonable success of hydrodynamics in describing not only heavy-ion, but also much smaller collision systems, triggered increased theoretical interest in understanding under which conditions hydrodynamics is applicable. In 2015, Israel-Stewart theory was analyzed in terms of hydrodynamic and non-hydrodynamic modes. Since the latter decay exponentially, the system relaxes to an attractor \cite{Heller:2015dha}, at which point the hydrodynamic modes dominate and hydrodynamics should be applicable. Many more works followed, exploring hydrodynamic behavior in many underlying theories. An excellent review on these topics can be found in \cite{Florkowski:2017olj}.

According to the fluctuation-dissipation theorem, viscous effects are tied to hydrodynamic fluctuations. A first exploration of their effects on the hydrodynamic evolution was performed in \cite{Kapusta:2011gt}. Their inclusion was also motivated by searches for the QCD critical point near which critical fluctuations are expected due to a diverging correlation length \cite{Hohenberg:1977ym, Gitterman:1978zz}. Several fluctuating hydrodynamic simulations with different approaches have since been developed \cite{Murase:2013tma, Young:2014pka, Stephanov:2017ghc, Murase:2019cwc, Sakai:2020pjw}. Generally, the effects of thermal fluctuations on the standard observables have been found to be mild, with some improvements found in describing anisotropic flow in ultra-central collisions \cite{Kuroki:2023ebq}.

Additional extensions of the standard viscous relativistic hydrodynamic framework introduced within the last ten years include spin- and magneto-hydrodynamics. To describe the evolution of spin and polarization in high-energy nuclear collisions, fluid equations were extended to describe the evolution of the polarization tensor \cite{Florkowski:2017ruc, Weickgenannt:2022zxs, Weickgenannt:2022qvh}. This is for example required in order to describe the measurable polarization of observed hadrons, such as $\Lambda$ hyperons and vector mesons \cite{STAR:2017ckg}. Magneto-hydrodynamics for heavy-ion collisions was developed in order to be able to study anomalous charge transport within the fluid in a magnetic field background; this is needed to make predictions for observables designed to discover the chiral magnetic effect \cite{Hattori:2017usa, Inghirami:2019mkc}.

The sensitivity of flow observables to the nuclear structure of projectile and target was also considered in the 2000s \cite{Heinz:2004ir}, with results from the STAR Collaboration demonstrating clear signals of the deformation of the uranium nucleus \cite{STAR:2015mki}. Studies of smaller collision systems such as O+O collisions, performed at RHIC and planned at LHC, as well as fixed target collisions performed and planned at LHCb using the SMOG and SMOG2 systems, will be able to probe this connection in lighter nuclei; given the more sophisticated theoretical understanding of their nuclear structure, and hence of the initial state geometry in collisions of smaller nuclei, these studies can yield better constraints on the fluid properties.

\begin{figure}[htbp]
\begin{center}
\includegraphics[width=13cm]{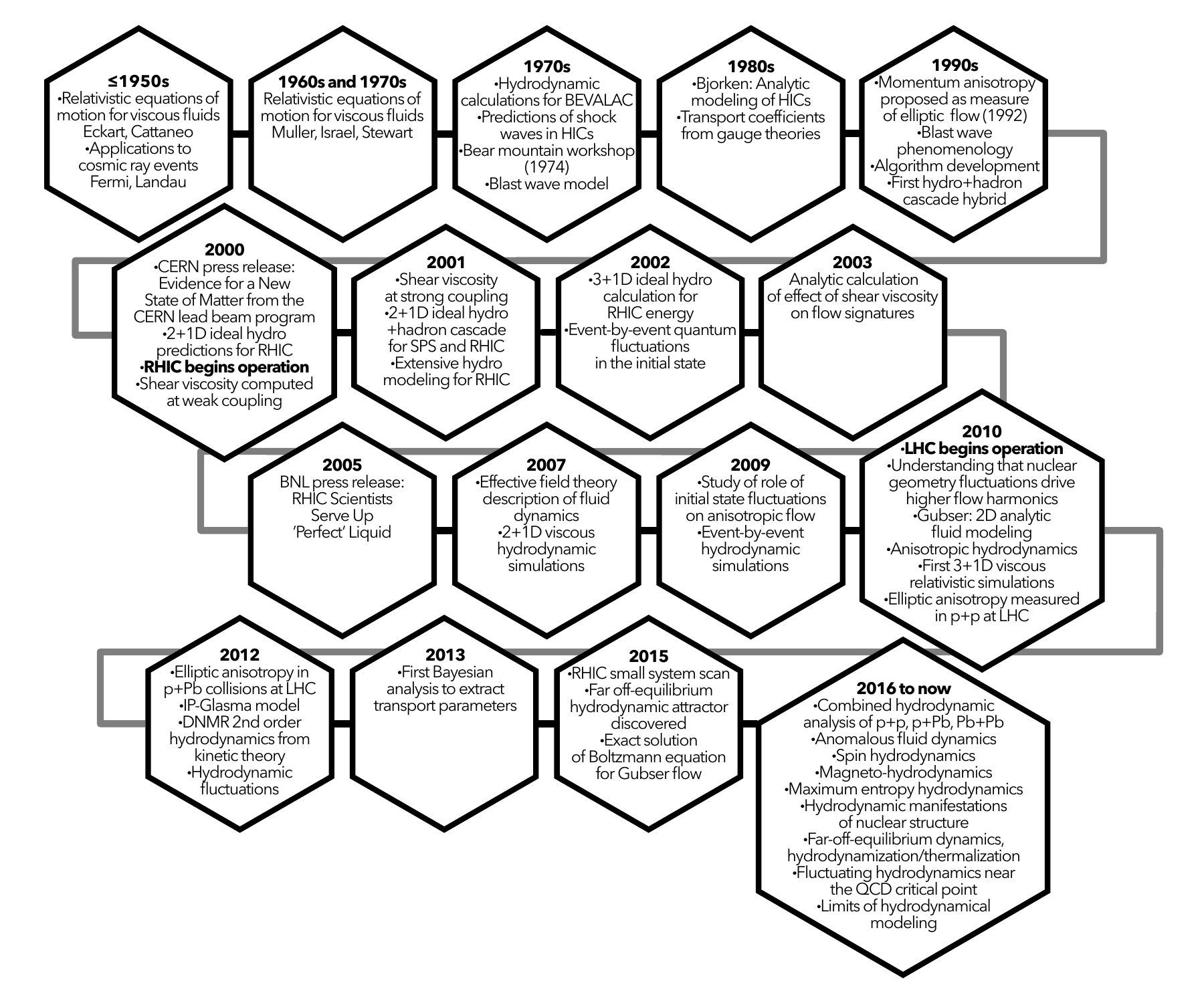}
\end{center}
\caption{A timeline of major events for the hydrodynamic description of the QGP.}
\label{fig:timeline}      
\end{figure}

A timeline with a concise summary of the major developments related to the hydrodynamic description of heavy-ion collisions is provided in Fig.\,\ref{fig:timeline}. 

\section{Relativistic hydrodynamics}
\label{sec:2}
The fluid dynamic description of high-energy heavy-ion collisions requires a relativistic formulation, because fluid velocities can reach values close to the speed of light. The relevant degrees of freedom are thus the energy and momentum density of the system, encoded in the stress-energy tensor $T^{\mu\nu}$, as well as the conserved charges, such as net-baryon, -strangeness, and -electric charge densities. First studies neglected any viscous effects and described the system as an ideal fluid, with the conservation equations taking the form
\begin{align}
    \partial_\mu T^{\mu\nu} =0\,,\\
    \partial_\mu J^\mu_i =0\,,
\end{align}
where $i \in \{B,S,Q\}$ labels the conserved charges. The stress-energy tensor is symmetric and constructed from the four-velocity of the fluid $u^\mu$, the metric $g^{\mu\nu}$, as well as the pressure and energy density in the local rest frame, $P$ and $\varepsilon$:
\begin{equation}\label{eq:Tmunu}
    T^{\mu\nu} = \varepsilon u^\mu u^\nu + P (u^\mu u^\nu -g^{\mu\nu})\,.
\end{equation}
The charge current is expressed as
\begin{equation}
    J^\mu_i = n_i u^\mu\,,
\end{equation}
where $n_i$ is the charge density. When considering one conserved current, this leaves six unknowns, $\varepsilon$, $P$, $u^\mu$ (3 unknowns since $u_\mu u^\mu=1$), and $n$, but so far we only wrote down 5 equations. The system needs to be closed by the equation of state $P(\varepsilon,n)$, which characterizes the underlying physical system and is an input into the hydrodynamic calculation. 

Viscous corrections enter as modifications to $T^{\mu\nu}$: 
\begin{equation}\label{eqn:tmunuvisc}
  T^{\mu\nu} \rightarrow T^{\mu\nu} + \Pi^{\mu\nu}\,,
\end{equation}
where $\Pi^{\mu\nu}$ can contain bulk, shear and heat flux corrections, the latter being typically eliminated by the choice of frame. The charge currents receive a diffusion contribution 
\begin{equation}\label{eqn:Jvisc}
  J^\mu_i \rightarrow J^\mu_i + \nu^\mu_i,
\end{equation}
where $\nu^\mu_i$ is the flow of net charge $i$ relative to $u^\mu$. Corrections are included systematically in gradients of $\varepsilon$, $P$, $u^\mu$, and $n_i$, leading to algebraic equations for the viscous corrections, the relativistic Navier-Stokes equations, at first order and to dynamic equations at higher order. As the Navier-Stokes equations do not have a dynamical description of the viscous stress tensor, it reacts instantaneously to the gradients in the system, leading in the relativistic case to problems with superluminal signal propagation; this causes instabilities in simulations.

Consequently, higher order equations are necessary\footnote{We comment on the possibility to construct a causal and stable first order theory below.} to study heavy-ion collisions within a viscous hydrodynamic framework. They introduce a relaxation time for the response of the viscous corrections in $T^{\mu\nu}$ to the temperature and flow gradients that drive the medium out of equilibrium, avoiding acausal behavior (at least at the linear response level).  Second order viscous hydrodynamic equations were derived by M\"uller and, independently, by Israel and Stewart \cite{ Muller:1967zza, Israel:1976, Israel:1979wp}. Using the second law of thermodynamics avoided the need to include knowledge about microscopic details of an underlying theory.
Derivations from kinetic theory are also possible, e.g. using the relativistic version of Grad's method of moments \cite{Grad:1949,Olson:1990rzl,Israel:1976}. 

The equations of motion obtained by Israel and Stewart have solutions that are known to be causal \cite{Hiscock:1983zz,Huang:2010sa} and linearly stable around equilibrium, as long as relaxation-time transport coefficients obey certain bounds. In realistic collisions one is not infinitesimally close to local equilibrium, and usually in the fully nonlinear regime. Necessary and sufficient conditions that ensure causality in the nonlinear regime of Israel-Stewart-like theories were derived in \cite{Bemfica:2019cop,Bemfica:2020xym}.
Further, complete second order equations were derived for conformal theories in \cite{Baier:2007ix}.

A set of equations, now often used in hydrodynamic simulations of heavy-ion collisions, was derived by Denicol, Niemi, Molnar, and Rischke (DNMR) \cite{Denicol:2012cn}. Here, all terms of the moment expansion are kept and exact evolution equations for the moments derived. Then, keeping only the equations of motion for the slowest microscopic modes and approximating faster modes by their asymptotic Navier-Stokes solutions, a combined expansion in Knudsen number and inverse Reynolds numbers is found, providing a systematic truncation with a well defined regime of applicability.\footnote{
{The DNMR scheme is {\it perturbative}, being based on the assumption that the deviations of the distribution function from local thermal equilibrium are generically small. In aHydro and ME Hydro (see below) this assumption is replaced by {\it non-perturbative} ansaetze for the distribution function designed to approximate the solution of the Boltzmann equation also in far-off-equilibrium situations.}} 
Recently, causal, stable, and locally well-posed first-order theories have been constructed \cite{Kovtun:2019hdm, Bemfica:2019knx}. Here, the important condition for stability is the right frame choice, with the Landau frame not providing stability, leading to the original problems with Navier-Stokes theory discussed above.

\section{Hydrodynamic modeling}
\label{sec:3}
\begin{figure}[htbp]
\begin{center}
\includegraphics[width=11cm]{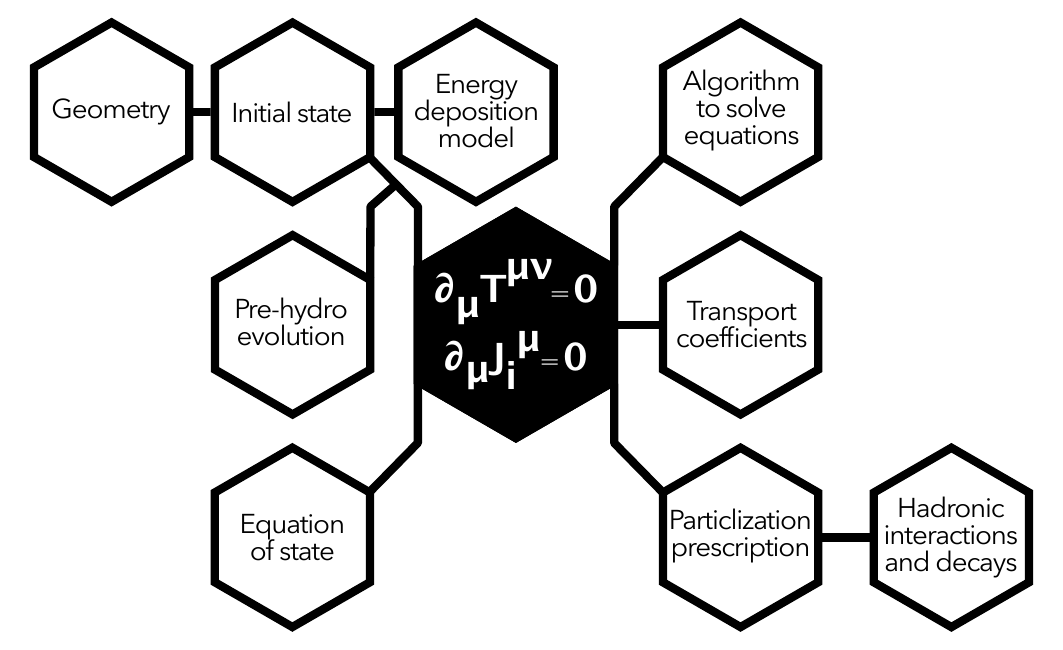}
\end{center}
\caption{Hydrodynamic models of heavy-ion collisions are centered around the fluid dynamic equations but involve many more components that can contain input from first principles calculations but often also need significant modeling.
}
\label{fig:model}      
\end{figure}

Naturally, the hydrodynamic equations are at the core of any hydrodynamic simulation of heavy-ion collisions. But such simulations have many more ingredients, requiring a significant amount of modeling  \cite{Heinz:2013th, Gale:2013da, Jeon:2015dfa, Romatschke:2017ejr}. We illustrate this in Fig.\,\ref{fig:model}, where we schematically sketch a hydrodynamic simulation framework.

\subsection{Numerical implementations}
A core ingredient is the numerical solver of the hydrodynamic equations. A variety of approaches have been taken in the field of heavy-ion phenomenology. Commonly used are finite volume methods, where one subdivides the spatial domain into grid cells and keeps track of an approximation to the average of the conserved quantity within these volumes. Such methods include the Lax-Friedrichs method \cite{leveque_numerical_1992}, the Nessyahu-Tadmor method \cite{NESSYAHU1990408}, the SHArp and Smooth Transport Algorithm (SHASTA) \cite{BORIS197338}, and the Kurganov-Tadmor (KT) method \cite{KURGANOV2000241}. The various algorithms have different approaches to reduce numerical diffusion, which is the main undesirable side effect of discretization. Also, Riemann solvers, first introduced by Godunov \cite{zbMATH03273813} have been used extensively to solve at least the ideal part of dissipative relativistic hydrodynamics of the QGP. Other algorithms include the Piecewise Parabolic Method (PPM) \cite{Hirano:2012yy}, and the forward in time, centered in space (FTCS) scheme \cite{Luzum:2008cw,Bozek:2011ua}. A rather different approach, not using a fixed spatial grid, is the Smoothed Particle Hydrodynamics (SPH) method \cite{Aguiar:2000hw, Noronha-Hostler:2013gga}. Such Lagrangian methods envision the fluid as composed of small objects that move in time and space.  One tracks the velocity and positions of those individual objects (or SPH ``particles''), and from there can reconstruct the entire motion of the fluid. 

\subsection{Initial state}
One of the most important ingredients of a hydrodynamic model for heavy-ion collisions is the description of the initial state. For most simulations, this involves an initial condition for the stress-energy tensor and the conserved currents as functions of the spatial location. The most important feature is the initial geometry, as it defines the pressure gradients that drive the hydrodynamic expansion.

Early initial state models were often based on the optical Glauber or Monte-Carlo (MC) Glauber models \cite{glauber1959,Miller:2007ri}. For high-energy collisions, one is typically working with thickness functions, the integrals of the nuclear density distributions of target and projectile over the longitudinal direction. The optical Glauber model applies a partial wave analysis and eikonal approximation to compute the inelastic cross section and expresses them in terms of thickness functions. It assumes independent linear trajectories for all nucleons and that the total phase shift function is the sum of the phase shift functions of each individual nucleon. 

The model does not provide information on particle production; additional assumptions are necessary, for example a wounded nucleon, binary collision, or two component model, which assume particle production or initial entropy density (or energy density) deposition to be proportional to the number of wounded nucleons and/or binary collisions, respectively.

Monte-Carlo Glauber models usually describe the collision on the nucleon level,\footnote{%
    Glauber models based on valence quarks as colliding degrees of freedom also exist (e.g. \cite{PHENIX:2013ehw, Schenke:2014zha, Welsh:2016siu}). Hydrodynamic triangular flow ($v_3$) in p+p collisions for example cannot be understood without accounting for fluctuating geometric structure on the subnucleonic level.}
employing the nucleon-nucleon cross section to determine wounded (or participant) nucleons on an event-by-event basis, which are then assumed to contribute to the initial energy or entropy densities. To account for multiplicity fluctuations in p+p collisions (where the number of wounded nucleons does not fluctuate), this contribution is assumed to fluctuate from event to event \cite{Chaudhuri:1988yh, McLerran:2015qxa}, with an amplitude typically sampled from a $\Gamma$-distribution  (see e.g. \cite{Shen:2014vra}). Thickness functions of wounded nucleons can be determined and used to define an initial energy density. Various functional forms have been used, with some modern models including a wide range of possibilities, such as \Trento\  \cite{Moreland:2014oya} and its 3-dimensional generalization \Trento-3D \cite{Soeder:2023vdn}. Comparison to experimental data then allows to constrain the parameters related to how energy is deposited.

Alternatively, there exist first principles calculations that provide the energy (and momentum, and sometimes charged current) deposition mechanism. As typical for analytic approaches in QCD, they rely on either weak or strong coupling approximations. 
In the limit of weak coupling, the so called Color Glass Condensate (CGC) effective theory \cite{McLerran:1993ni, Iancu:2003xm, Gelis:2010nm, Blaizot:2016qgz} can be used to describe colliding systems of high gluon density. This is done, for example, in \cite{Hirano:2004en} and in the IP-Glasma model that, in conjunction with viscous fluid dynamics, has proven rather successful in describing many features of heavy-ion collisions \cite{Schenke:2012wb, Gale:2012rq}. It relies on the MC-sampling of nucleons and subnucleonic degrees of freedom, including hot spots and color charges, which form incoming color currents, which in turn source the incoming dynamic gluon fields as dictated by the SU(3) Yang-Mills equations. From these one can determine the gluon fields generated in the collision and their stress-energy tensor. In collisions with proton projectiles, IP-Glasma model calculations require a nucleon substructure to describe the anisotropic flow \cite{Schenke:2014zha}. Including substructure introduces more parameters, which can be constrained using e+p scattering data \cite{Mantysaari:2016ykx,Mantysaari:2017cni}.

It is also possible to use collinear factorization of perturbative QCD (pQCD) to compute the production of the dominant few-GeV gluons, a.k.a. ``minijets'', including gluon-saturation effects; one example is the EKRT model, named after the authors \cite{Eskola:1999fc}. Alternatively, one can work in the limit of strong coupling, making use of the duality between a conformal field theory (CFT) and gravity in Anti-de-Sitter (AdS) space (dubbed AdS/CFT), which was established in \cite{Maldacena:1997re}, in order to determine the initial energy momentum tensor produced in a nuclear collision. We note that strong coupling calculations \cite{Chesler:2013lia, Gubser:2014qua, Chesler:2015lsa} typically have led to narrower rapidity distributions of emitted particles than what has been found experimentally \cite{vanderSchee:2015rta}.

Other initial state models that have been employed in hydrodynamic simulations of heavy-ion collisions range from PYTHIA \cite{Sjostrand:2007gs} based models, such as the dynamical core–corona initialization (DCCI) model \cite{Kanakubo:2019ogh}, to HIJING \cite{Wang:1991hta} and (HIJING based) AMPT \cite{Lin:2004en} models, to the initial stages of hadron transport codes, such as UrQMD \cite{Bass:1998ca, Bleicher:1999xi} or SMASH \cite{Weil:2016zrk}, which typically use a string based description of particle production. For low energy collisions, dynamical initial state models have emerged. They use a source term in the hydrodynamic equations to deposit energy over a finite amount of time, while the nuclei are passing through each other. One such model is based on UrQMD \cite{Karpenko:2015xea}, another extends an MC-Glauber model to three dimensions and uses string deceleration between colliding nucleons or valence quarks to dynamically determine the deposited energy and momentum \cite{Shen:2017bsr, Shen:2022oyg}.

Extensions to three dimensions have also been developed within the IP-Glasma model framework, by invoking rapidity evolution within the CGC \cite{Schenke:2016ksl, Schenke:2022mjv, McDonald:2023qwc}. Alternatively, a finite nuclear thickness in the longitudinal direction can also been used to determine a non-boost-invariant glasma initial condition \cite{Ipp:2017lho}.

%%%%%%%%%%%%%%%%%%%%%%%%%%%%%%%%%%%%%%%%%%
\subsection{Early time dynamics and thermalization}
\label{sec:therm}
%%%%%%%%%%%%%%%%%%%%%%%%%%%%%%%%%%%%%%%%%%
%

According to text book knowledge, the underlying assumption for the applicability of hydrodynamics is that of approximate local thermal equilibrium.\footnote{%
    A medium in perfect local equilibrium evolves as an ideal (inviscid) fluid. As long as the deviations from perfect local equilibrium remain small, viscous corrections to $T^{\mu\nu}$ and $J^\mu$ can be handled within standard dissipative fluid dynamics. The possibility of describing far-off-equilibrium media with suitably structured far-off-equilibrium macroscopic evolution equations remains an active research area.}
Does the system reach thermal equilibrium fast enough, so we can employ hydrodynamics at times as early as a fraction of $1\,{\rm fm}/c$? This question has been the focus of a significant amount of research, and we can only briefly outline the progress made towards answering it.

Already in the year 2000, the bottom-up picture of thermalization \cite{Baier:2000sb} was established. This weak-coupling picture is based on an initially over-occupied and momentum-anisotropic gluon-dominated state called glasma (as that in the IP-Glasma model) and proceeds in stages: 1) Expansion and collisional broadening compete and lead to a momentum diffusion in the longitudinal direction while decreasing the gluon occupation number to order one. 2) ``Soft'' gluons are emitted via medium induced collinear radiation and eventually overwhelm in number the primary ``hard'' gluons (defined as having momenta of order $Q_s$, the saturation scale). The soft gluons quickly equilibrate and form a thermal bath, but the higher momentum gluons still carry the dominant fraction of the total energy. 3) The thermal bath draws energy from the harder gluons. The system finally thermalizes when the energy in the soft and hard components becomes comparable.

The early stages of this picture were verified numerically within classical statistical simulations \cite{Berges:2013eia, Berges:2014yta}, with one difference being that at very early times Weibel-like plasma instabilities also play a role, affecting the isotropy of the early overoccupied plasma. The possible importance of plasma instabilities for thermalization in heavy-ion collisions was already pointed out in the 1990s \cite{Mrowczynski:1993qm}, and a lot of subsequent work was dedicated to understanding their role \cite{Mrowczynski:2016etf}. The classical statistical simulations indicate that plasma instabilities only dominate at the earliest times after which a universal scaling behavior emerges \cite{Berges:2014bba}, which is described in terms of non-thermal attractor solutions \cite{Berges:2013eia, Berges:2020fwq}. As the occupation number decreases, these numerical solutions in the classical regime are no longer valid and a quantum description is required. The later stages of the bottom-up scenario discussed above provide such a description.

The studies outlined above are valid in the limit of weak coupling. To apply them to realistic situations of heavy-ion collisions, extrapolations to realistic couplings are necessary. This is typically done within QCD kinetic theory. Based on a non-equilibrium linear response formalism, a concrete realization of an effective kinetic theory stage bridging the gap between the initial state and hydrodynamics was introduced in \cite{Kurkela:2018wud}.

To describe real-time phenomena in strongly coupled (3+1)-dimensional quantum field theories one has to resort to holography \cite{Maldacena:1997re}, which is based on the notion of a correspondence of the gauge theory degrees of freedom to higher dimensional geometries, which arise as solutions of Einstein’s equations with a negative cosmological constant and appropriate matter fields. As one is interested in time-dependent states in Minkowski spacetime that model the dynamics of heavy-ion collisions, one needs to study expectation values of the energy-momentum tensor by solving the equations of motion of the gravity action as an initial value problem, using numerical relativity techniques (see \cite{Berges:2020fwq} for more details). 

In such studies it was found that low-order hydrodynamic constitutive relations become applicable at strong coupling after a time of order of the inverse temperature. In this regime the pressure anisotropy in the system is still sizable. Because of this, the term ``hydrodynamization'' was coined in \cite{Casalderrey-Solana:2011dxg} to distinguish the applicability of viscous hydrodynamic constitutive relations from local thermalization.   

Often hydrodynamic simulations have ignored the question of how thermalization is achieved dynamically, and assume a close to thermal initial state at the initial time. This brings with it some problems: for example, the initial condition provided by the IP-Glasma model, with zero longitudinal pressure and positive transverse pressure, is far from equilibrium. Such deviations from equilibrium can be absorbed into the initial viscous stress tensor but they are typically far from small corrections. Improvements can be achieved by the addition of an intermediate transport stage preceding the hydrodynamic stage \cite{Kurkela:2018vqr, Kurkela:2018wud, Kurkela:2019kip, Ambrus:2022koq, Ambrus:2022qya}. In (0+1)-dimensional systems undergoing Bjorken expansion, anisotropic hydrodynamics (aHydro) has been shown \cite{Florkowski:2010cf, Martinez:2010sc} to accurately describe this far-off-equilibrium transport stage, for media consisting of both massless or massive particles. 

The idea of aHydro was recently generalized into Maximum Entropy Hydrodynamics (ME Hydro) \cite{Chattopadhyay:2023hpd}, by replacing the anisotropic distribution function used in deriving aHydro with a maximum entropy (least biased) distribution which is the most likely distribution underlying the energy-momentum tensor when the latter features large viscous corrections {\it in any direction}. aHydro and ME Hydro were found to perform very well in systems undergoing Bjorken and Gubser flow \cite{Chattopadhyay:2023hpd}, but a test in more general flow situations such as those encountered in the later evolution stages of heavy-ion collisions is still outstanding. 

%%%%%%%%%%%%%%%%%%%%%%%%%%%%%%%%%%%
\subsection{Equation of state}
%%%%%%%%%%%%%%%%%%%%%%%%%%%%%%%%%%%
%
The equation of state contains information on the underlying microscopic theory, in our case QCD. In the early stages of the phenomenological description of heavy-ion collisions, the equation of state at (close to) zero baryon chemical potential $\mu_B$ was not known from first principles.
Therefore, besides an ideal gas equation of state, an equation of state with a first-order phase transition based on the MIT bag model \cite{Chodos:1974je, Lee:1986mn, Heinz:1987sj, Sollfrank:1996hd, Hung:1997du} was widely used. As lattice QCD calculations at zero baryon chemical potential became available, lattice based equations of state \cite{Ratti:2022qgf} were employed in hydrodynamic simulations. They predict a smooth cross over, as opposed to the first-order phase transition of earlier model equations of state. At low temperatures they are matched to hadron resonance gas equations of state. Initially, there were problems matching the lattice to hadron resonance gas results \cite{Huovinen:2009yb}, but this had to do with approximations (e.g.~the use of unphysical quark masses) used in early lattice calculations and has since been resolved. The now available lattice QCD based equations of state represent solid first principles results at zero $\mu_B$ \cite{HotQCD:2014kol, Borsanyi:2013bia, Moreland:2015dvc}.

Extracting the equation of state from the lattice at finite $\mu_B$ is challenging, because of the sign problem \cite{deForcrand:2009zkb}. The inclusion of the chemical potential makes the fermion determinant complex, leading to oscillations in the path integral weight. This complicates numerical simulations, as standard reweighing techniques become inefficient and unreliable. Various techniques have been developed to circumvent this problem, for example Taylor expansion \cite{Gavai:2001fr, Allton:2002zi}, the imaginary chemical potential method \cite{deForcrand:2002hgr, DElia:2002tig, Gunther:2016vcp}, Lefschetz thimble decomposition \cite{Pham:1983, Witten:2010cx}, or the complex Langevin method \cite{Parisi:1984cs, Ambjorn:1985iw}.

The latest lattice QCD results at vanishing chemical potentials \cite{Bazavov:2014pvz, Bazavov:2012jq, Ding:2015fca, Bazavov:2017dus, HotQCD:2018pds, Borsanyi:2020fev} allow for the construction of a multi-dimensional equation of state in temperature, baryon-, strangeness-, and electric charge chemical potentials \cite{Monnai:2019hkn, Noronha-Hostler:2019ayj, Monnai:2024pvy}. Such constructions are important for the study of charge transport in heavy-ion collisions, particularly at lower beam energies, where the net-baryon charge may be significant. Further, if one is interested in fluctuation driven observables, strangeness and charge fluctuations also need to be considered, making the use of a multi-dimensional equation of state necessary. 

While lattice calculations have so far not seen any evidence of this, model calculations at large $\mu_B$ suggest the possible existence of a critical point at some finite temperature and chemical potential where the crossover transition turns into a first-order phase transition \cite{Fukushima:2010bq, Asakawa:1989bq}. Beam energy scan (BES) programs at RHIC and the CERN SPS, along with future efforts at FAIR, NICA, and J-PARC, aim to explore finite-density QCD matter and determine experimentally if such a QCD critical point exists. Equations of state that include a critical point have been modeled \cite{Parotto:2018pwx, Karthein:2021nxe} to facilitate studying its effects in hydrodynamic simulations.

Recently, efforts have emerged to collect and combine knowledge from both heavy-ion collisions and neutron star mergers to map out the phase diagram of strongly interacting matter \cite{Dexheimer:2020zzs}.

%%%%%%%%%%%%%%%%%%%%%%%%%%%%%%%%%%%%%%%%
\subsection{Transport coefficients}
\label{sec:TC}
%%%%%%%%%%%%%%%%%%%%%%%%%%%%%%%%%%%%%%%%
Another input into the hydrodynamic calculations are the transport coefficients {of the fluid medium}, such as the shear and bulk viscosity to entropy density ratios, $\eta/s$ and $\zeta/s$, and the corresponding relaxation times, as well as other higher order coefficients. 

Determining the transport coefficients and their temperature and chemical potential dependence from QCD is extremely challenging. Consequently, a typical approach is to make some assumptions about the form of the higher order coefficients and extract the values of $\eta/s$ and $\zeta/s$ (and their temperature and chemical potential dependence) from comparisons of hydrodynamical model predictions with experimental data.

Alternatively, transport coefficients can be computed from first principles, at least in certain limits. The usual approaches include perturbative QCD, AdS/CFT, lattice QCD, and effective field theories.

Generally, transport coefficients can be computed in quantum field theory using Kubo formulas, which relate the coefficients to equilibrium correlation functions of the corresponding currents. The Kubo formulas can be derived within linear response theory, using the fluctuation dissipation theorem. 

At very high temperatures, where the coupling constant becomes small, transport coefficients can be computed systematically using kinetic theory based on the Boltzmann equation or its quantum extensions. The QGP shear viscosity, electrical conductivity, and flavor diffusion constants were computed within this limit in \cite{Arnold:2000dr}. Bulk viscosity requires a significantly different, and more complicated, analysis, and was computed later in \cite{Arnold:2006fz}. These results were done at leading (or `leading-log') order in the QCD coupling constant $\alpha_s$. Next-to-leading order calculations for the shear viscosity were performed more recently in \cite{Ghiglieri:2018dib}, and corrections found to be large, reducing $\eta/s$ by more than a factor of 3 at physically relevant couplings. Expressed differently, one finds that the perturbative series for the shear viscosity only converges at temperatures well above 100 GeV. 
The leading order shear viscosity of QCD was computed using the resummed 3PI effective action in \cite{Carrington:2009xf}. Quasiparticle models \cite{Bluhm:2010qf} and parton cascades \cite{Xu:2007ns, Wesp:2011yy} with tunable coupling have also been used to extract transport coefficients.

In the opposite, extremely strong coupling limit, calculations were performed using holography. The first such calculation was performed in \cite{Policastro:2001yc}, providing the value $\eta/s=1/(4\pi)$ for a conformal $\mathcal{N}{\,=\,}4$ supersymmetric Yang-Mills plasma. Subsequent studies within a wide range of theories led to the conjecture that this value serves as a lower bound for $\eta/s$ \cite{Buchel:2003tz, Kovtun:2004de, Iqbal:2008by}; however, counter examples have since been found. The bulk viscosity has been calculated using holography for non-conformal gauge theories \cite{Benincasa:2005iv}, and the form of higher order transport coefficients can also be determined using holography \cite{Baier:2007ix, ALICE:2012eyl}. We refer the reader to \cite{Casalderrey-Solana:2011dxg, Brambilla:2014jmp, Jarvinen:2021jbd} for more details on the application of holography to computing transport coefficients.

To avoid taking extreme limits of weak or strong coupling, one can attempt computing transport coefficients on the lattice. The lattice can provide correlation functions in Euclidean time, but the Kubo formulas used to compute transport coefficients are written in terms of Minkowski time. The correlation function from the lattice can be seen as one in imaginary time, but a direct relation to the desired correlator exists. However, the actual reconstruction of the correlation function that appears in the Kubo formula, in particular in the limit of vanishing frequency, is challenging if not impossible, as very distinct Minkowski functions can arise from nearly identical Euclidean continuations \cite{Moore:2020pfu}. Recent results for pure glue shear viscosity from the lattice can be found in \cite{Altenkort:2022yhb}.

Functional renormalization group computations have also extracted the temperature dependent shear viscosity to entropy density ratio by using non-perturbative techniques in the intermediate temperature regime around $T_c$ \cite{Haas:2013hpa, Christiansen:2014ypa}.

%%%%%%%%%%%%%%%%%%%%%%%%%%%%%
\subsection{Late stages and hadronic freeze-out}
%%%%%%%%%%%%%%%%%%%%%%%%%%%%%

Hydrodynamic simulations describe a (typically one-component) fluid, but heavy-ion experiments measure produced particles moving through the detector. Consequently, any hydrodynamic simulation of heavy-ion collisions needs a prescription for converting the fluid into final particle distributions (``particlization''). Naturally, as the hydrodynamic medium expands, it becomes increasingly dilute, and the strongly interacting description ceases to be warranted; a description in terms of a dilute gas becomes more appropriate. With an underlying microscopic theory in mind, one can argue that the transition should occur when the microscopic interaction rate drops below the macroscopic expansion rate \cite{Lee:1988rd, Schnedermann:1992ra}. In practice, one often uses a specific energy density or temperature as the criterion for determining the surface beyond which the system is either considered completely frozen out or more accurately described as a dilute gas of hadrons and hadronic resonances that can have some residual interactions and possibly decay. 

Numerical simulations require a particlization surface finder, which determines the hypersurface of e.g.~constant energy density or temperature in space and time. These algorithms are rather complex as, for 3+1D simulations, they have to find a 3D surface inside a 4D volume (see e.g.~\cite{Schenke:2010nt, Huovinen:2012is}). For fluids  featuring large inhomogeneities in energy density the particlization hypersurface can have multiple simply-connected pieces.

To compute particle production one assumes that at particlization 
every infinitesimal part of the hyper-surface behaves like a simple black body source of particles. Thermal spectra of all particles are then given by the Cooper-Frye \cite{Cooper:1974mv} formula, modified by viscous corrections in case of dissipative hydrodynamics. The particles are boosted by the flow velocity of the cell from which they emerge, which imprints the (anisotropic) flow of the fluid onto the particle distributions.

In the early calculations, particle spectra as functions of transverse momentum (and possibly rapidity) were computed and resonance decays performed \cite{Sollfrank:1990qz, Mazeliauskas:2018irt} directly on the particlization surface (at that time called the ``freeze-out surface''). Performing consecutive decays, starting with the highest mass resonances, allows to compute the entire decay chain.
Measured are the energy and momentum distributions of stable charged hadrons (which leave ionization tracks in the detector), i.e. charged pions and kaons as well as protons and antiprotons. These include both directly emitted particles (described by flow-boosted thermal spectra with viscous corrections) and resonance decay products. Such a prescription neglects rescatterings (see below), as well as multi-particle correlations arising from the decay kinematics, which contribute to so-called ``non-flow'' correlations. Experiments try to eliminate the latter when aiming to measure flow observables. 

To include hadronic interactions after converting the fluid to particles, one samples the (viscosity-modified) Cooper-Frye spectra of all (charged and neutral, stable and unstable) hadrons emitted from the particlization surface event by event and feeds them into hadronic transport codes such as UrQMD \cite{Bass:1998ca, Bleicher:1999xi}, JAM \cite{Hirano:2005xf, Hirano:2007ei}, or SMASH \cite{Weil:2016zrk}. These codes will take care of any subsequent scatterings and all particle decays, leading to kinetic freeze-out naturally and dynamically (for more information see e.g.~\cite{Shen:2014vra}). To increase statistics for the calculated final particle distributions, one often generates from a single hydrodynamical event multiple hadronic final states by sampling the spectra of its particlization surface repeatedly and processing the sampled hadronic events independently through the hadronic transport code.

For many observables the assumption of a grand canonical system and the above sampling method is fine. However, when interested in correlations and fluctuations of conserved quantities, one should conserve energy-momentum and charges locally in every sampled event. This requires a local micro-canonical sampler. Otherwise correlations originating from conservation laws are lost. In principle, one should fulfill conservation laws exactly in every surface element. However, the number of produced particles is usually not large enough to make this feasible (often, the average number of particles to emerge from one cell is smaller than one). Instead, one can group surface elements into larger surface patches and impose conservation laws on those patches. New developments include micro-canonical sampling in such local patches \cite{Oliinychenko:2019zfk, Oliinychenko:2020cmr}.

Already in the 1990s it was established that chemical freeze-out (i.e. the decoupling of final particle abundances) precedes kinetic freeze-out (where the momentum spectra of emitted hadrons decouple) \cite{Schnedermann:1992ra, Heinz:1999kb, Braun-Munzinger:1999hun}. In principle, the particlization surface for the conversion of hydrodynamic fields to hadrons need not agree with either of these two surfaces; instead, it should be placed within a space-time region where the macroscopic hydrodynamic description and the microscopic hadron cascade picture both provide reasonably accurate dynamical models simultaneously. It is now broadly agreed that chemical freeze-out basically coincides with the quark-hadron phase transition \cite{Andronic:2017pug} such that a description in terms of a hadronic cascade can only be contemplated below the transition temperature. At particlization the matter is therefore still in approximate thermal (kinetic) equilibrium, for viscous hydrodynamics to still be valid, but already out of chemical equilibrium.\footnote{%
Unless particlization is enforced right at the phase transition where, however, the hadron cascade approximation may not yet be applicable.
  }
In that case a ``Partial Chemical Equilibrium'' (PCE) equation of state \cite{Gerber:1988tt} should be used for the hydrodynamic evolution between $T_c$ and particlization, and on the particlization surface the Cooper-Frye spectra must be sampled with corresponding PCE chemical potentials \cite{Hirano:2002ds, Huovinen:2007xh}.

%%%%%%%%%%%%%%%%%%%%%%%%%%%%%%%%%%%%%%%%%%%%%%
\section{Insights from hydrodynamic modeling of heavy-ion collisions}
\label{sec:4}
%%%%%%%%%%%%%%%%%%%%%%%%%%%%%%%%%%%%%%%%%%%%%%

Having discussed the basic ingredients of hydrodynamic models, we present in this section a concise summary of the key insights and developments driven by the application of relativistic hydrodynamics to heavy-ion collisions.

~\\
{\bf Discovery of the quark-gluon plasma}

The evidence for the creation of a strongly interacting quark-gluon plasma in high energy heavy-ion collisions is striking. Hydrodynamic models have been successful in predicting and describing consistently a wide range of observables. 

Hydrodynamic models with fluctuating initial conditions are able to reproduce multiplicity distributions, particle spectra as functions of $p_T$ and rapidity (in case of 3+1D simulations), as well as many observables sensitive to the initial geometry and its fluctuations, ranging from relatively simple $v_n$ coefficients to observables constructed from complex many-particle correlations.

\begin{figure}[tb]
  \centering
    \includegraphics[width=0.48\textwidth]{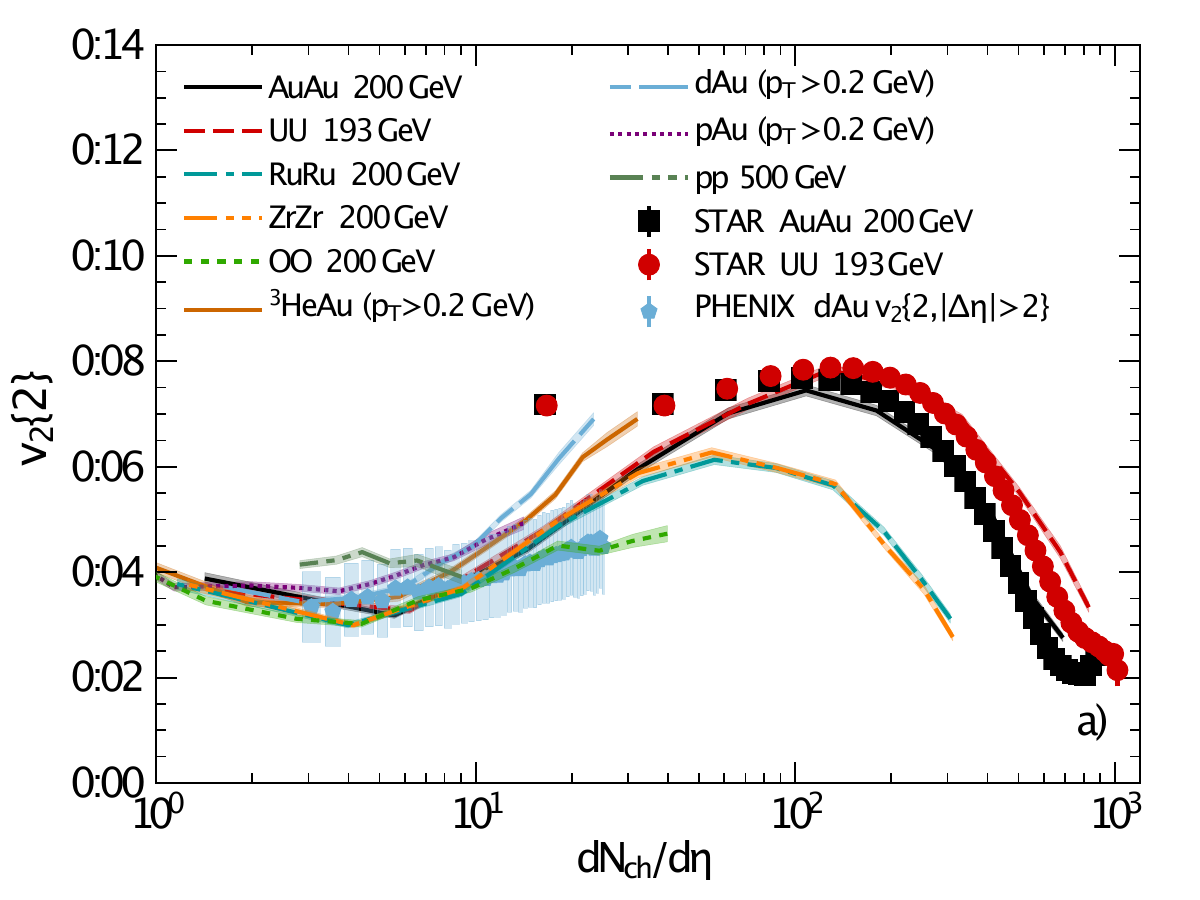}
    \includegraphics[width=0.48\textwidth]{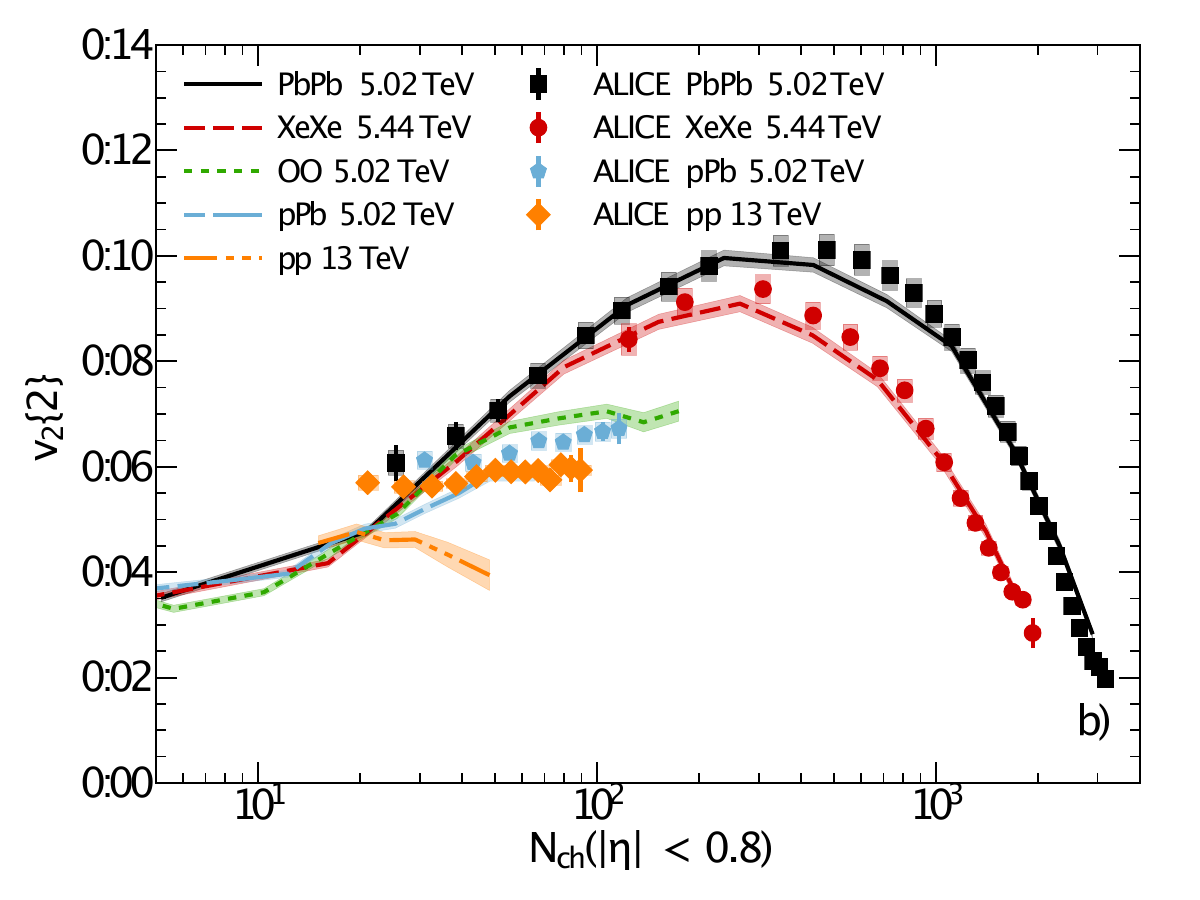}\\
  \caption{Anisotropy coefficients $v_2\{2\}$ (the $\{2\}$ indicating they are measured using 2-particle correlations) for charged hadrons as functions of multiplicity in various collision systems at (a) RHIC, compared to experimental data from the STAR \cite{STAR:2015mki, STAR:2019zaf} and PHENIX  \cite{PHENIX:2017xrm} Collaborations, and (b) the LHC, compared to experimental data from the ALICE Collaboration \cite{ALICE:2019zfl}. Calculations are from the IP-Glasma+MUSIC+UrQMD model \cite{Schenke:2020mbo}.
  \label{fig:v22}}
\end{figure}

As a striking, but in no way comprehensive, example of how well hydrodynamic models can reproduce experimentally measured flow harmonics over a wide range of systems, centralities, and collision energy, we show in Fig.~\ref{fig:v22} a comparison of IP-Glasma+MUSIC+UrQMD calculations \cite{Schenke:2020mbo} with experimental data. While the large systems are quantitatively well described for most (except the most peripheral) centralities (here represented by the charged particle multiplicities $dN_{\rm ch}/d\eta$ and $N_{\rm ch}$, respectively), some discrepancies appear for smaller systems, where the initial state is not as well under control and hydrodynamics is pushed to the limits of its applicability, with viscous corrections becoming large. In particular, quantitative agreement is not yet achieved for p+p collisions.

While the strength of the observed anisotropic flow of inclusive charged hadrons by itself already indicates the early build-up of flow during the QGP phase, there is further evidence that the flow is created within deconfined matter:
Studying the anisotropic flows of identified particles, one finds a characteristic grouping of the flow coefficients by particle type, namely mesons and baryons 
\cite{STAR:2022tfp, ALICE:2018yph}. Scaling by the number of quarks (2 or 3, respectively), one finds that the $p_T$-dependent flow coefficients, $v_n(p_T)$, of all particle species collapse approximately onto a universal curve, strongly suggesting (see e.g. \cite{Fries:2003vb, Muller:2012zq}) that the flow is carried by the valence quarks and generated early during the deconfined partonic phase.

~\\
{\bf Constraints on transport properties and the nuclear equation of state}

Higher order flow harmonics $v_n$, with $n{\,>\,}2$, are increasingly sensitive to viscous effects \cite{Staig:2010pn, Schenke:2011bn}. Their combined study thus helped to put more stringent limits on the transport coefficients. Furthermore, many rather complex observables, including symmetric cumulants, $v_n$-$\langle p_T \rangle$ correlations, and non-linear flow mode-coupling coefficients, have been proposed and measured \cite{ALICE:2022wpn}. Their detailed study allows to further disentangle initial state properties from transport properties of the probed fluid.  

\begin{figure}[htbp]
\begin{center}
\includegraphics[width=11cm]{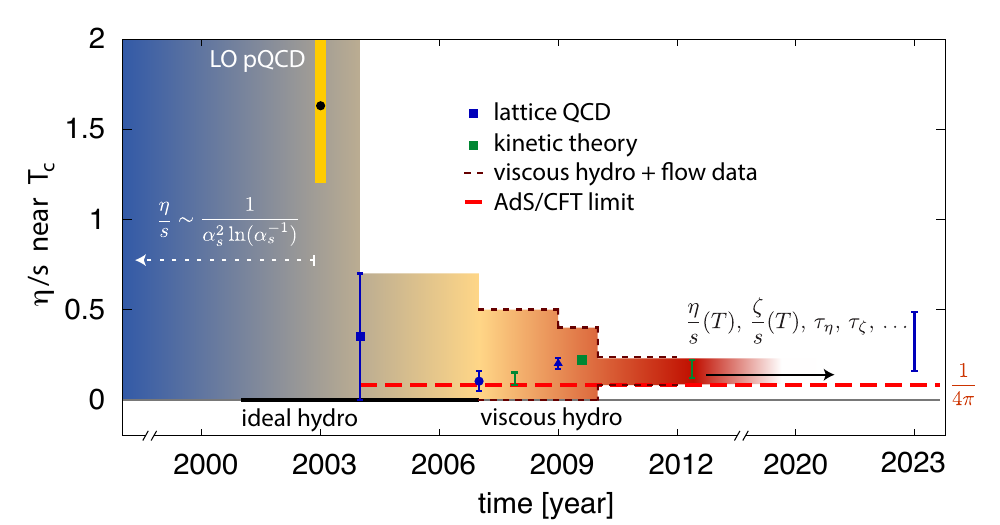}
\end{center}
\caption{The increasing precision of one key quantity, the shear viscosity to entropy density ratio $\eta/s$ near its minimal value, is illustrated. Shown are results from perturbative QCD calculations at leading logarithmic order \cite{Hosoya:1983xm, vonOertzen:1990ad, Thoma:1991em, Arnold:2000dr} (shown formula), complete leading order result \cite{Arnold:2003zc} (band from varying the scale by 20\%), Anti-deSitter gravity/Conformal Field Theory (AdS/CFT) correspondence \cite{Kovtun:2004de}, lattice QCD -- pure glue at $\sim 1.6\,T_c$, $1.24\,T_c$, $1.58\,T_c$, and $1.5\,T_c$, respectively \cite{Nakamura:2004sy, Meyer:2007ic, Meyer:2009jp, Altenkort:2022yhb}, ideal hydrodynamics \cite{Kolb:2000sd, Kolb:2000fha}, perturbative QCD/kinetic theory \cite{Xu:2007jv, Chen:2009sm, Ozvenchuk:2012kh}, and viscous hydrodynamics constrained by flow measurements \cite{Romatschke:2007mq, Luzum:2008cw, Song:2008hj, Song:2010mg}. For the last decade efforts focused on constraining the temperature dependence of the transport coefficients.}
\label{fig:etaOvers}      
\end{figure}

With the help of such measurements and improved event-by-event hydrodynamic models, the quantitative extraction of transport parameters from the experimental data has made great progress. As an example, we show in Fig.~\ref{fig:etaOvers} how significantly viscous hydrodynamic calculations contributed to the evolution of our knowledge of $\eta/s$ near the transition temperature $T_c$. Along with constraints from comparing hydrodynamic calculations with experimental data, we show purely theoretical extractions of $\eta/s$ from perturbative QCD, lattice QCD, kinetic theory, and the AdS/CFT correspondence.

\begin{figure}[htb]
    \centering
   \includegraphics[width=0.8\linewidth]{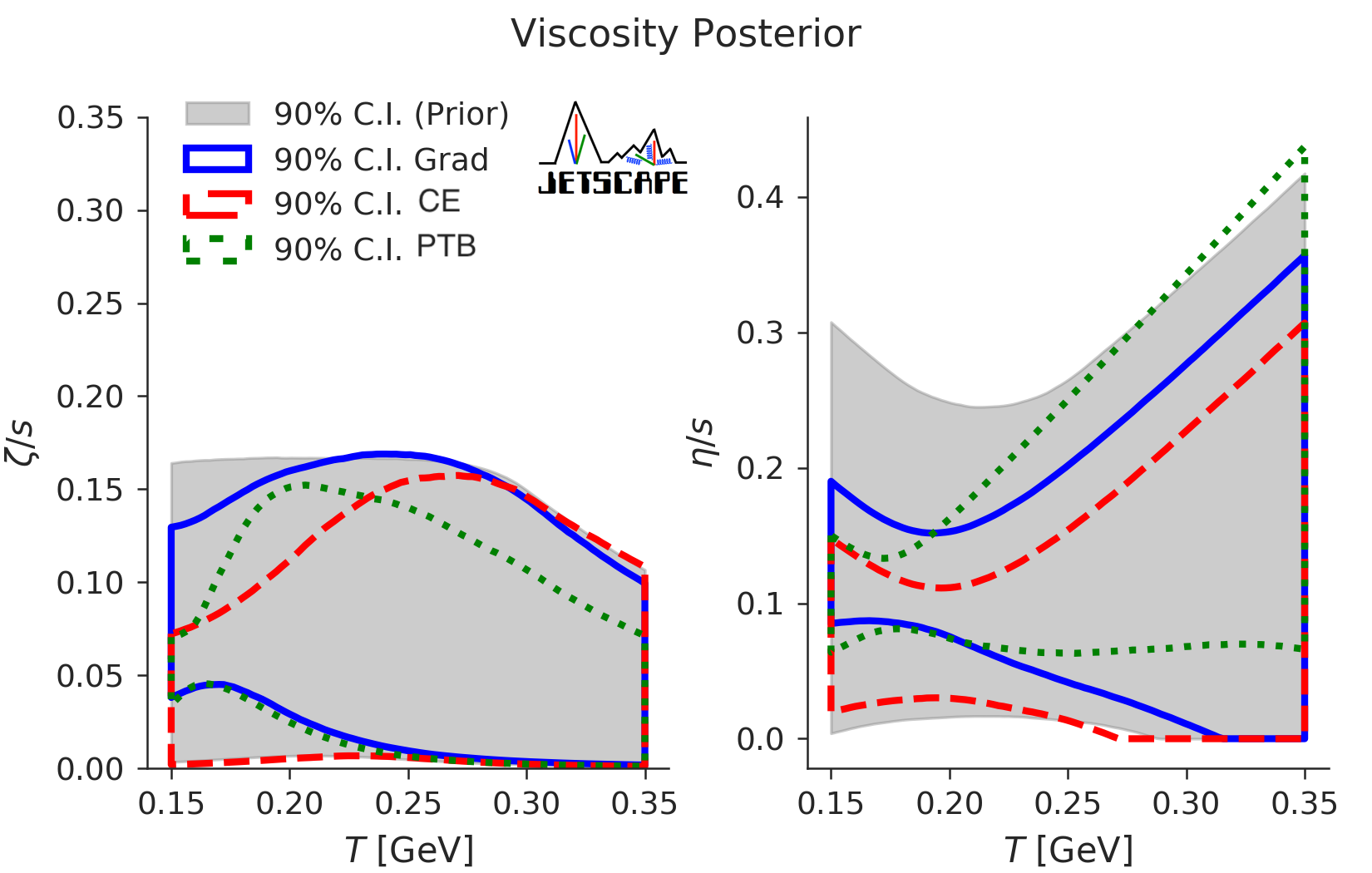}
    \caption{The 90\% credibility intervals for the prior (gray shaded area) and for the posteriors (colored outlines) of the specific bulk (left) and shear (right) viscosities, for three different models that describe the viscous correction to the particle distribution at particlization: Grad 14-moment approximation (blue), Chapman-Enskog (CE, red) and Pratt-Torrieri-Bernhard (PTB, green). 
    Note that different particlization prescriptions lead to different, but statistically mutually consistent posterior credibility intervals for the inferred viscosities, indicating a need for better quantification of the theoretical modeling uncertainties in Bayesian inference. Figure from \cite{JETSCAPE:2020mzn} (Creative Commons Attribution 4.0).}
    \label{fig:Bayesian}
\end{figure}

As the hydrodynamic model has (in addition to the QGP transport coefficients) many other parameters, in particular in the modeling of the initial state and the particlization of the QGP fluid, knowledge extraction from the experimental measurements is now more and more routinely done using Bayesian inference techniques \cite{Novak:2013bqa, Pratt:2015zsa, Bernhard:2015hxa, Bernhard:2019bmu, JETSCAPE:2020mzn, Nijs:2020ors, Liyanage:2023nds, Paquet:2023rfd, Soeder:2023vdn, Heinz:2023kzr, Hirvonen:2023lqy, JETSCAPE:2024cqe, Heffernan:2023utr, Heffernan:2023gye, Falcao:2024zkw}. 
Here, the inputs are experimental data and prior probability distributions for the model parameters that reflect our physics knowledge before the analysis (from theory and/or experimental data from other areas of physics, or from earlier, less precise or comprehensive measurements). Using Bayes'\ theorem, one obtains as output a posterior probability distribution for the model parameters and their correlations. This allows to constrain quantities such as $(\eta/s)(T)$ and $(\zeta/s)(T)$, along with estimates of their uncertainty. An example extraction of these quantities from a Bayesian analysis is shown in Fig.~\ref{fig:Bayesian}.

Bayesian inference can also be used to constrain the equation of state \cite{Pratt:2015zsa}, which is particularly relevant in the regime of non-zero baryon chemical potential where first principles calculations are not yet available \cite{Gong:2024lhq}. Studies of heavy-ion collisions at different collision energies using hydrodynamics are expected to contribute along with data from neutron star mergers to constrain the nuclear equation of state over a wide range of temperatures and densities \cite{Dexheimer:2020zzs}. 

To be able to run the many training points (choices of parameters) that emulators need within Bayesian analysis frameworks \cite{Paquet:2023rfd}, especially within event-by-event descriptions, recently machine learning tools have emerged \cite{Pang:2016vdc, Huang:2018fzn, Mallick:2022alr, Hirvonen:2023lqy, Heffernan:2023gye}. After being trained on hydrodynamic results themselves, these allow much faster prediction of final state quantities given an initial state. Such methods are likely to increase in importance for handling complex 3+1D event by event hydrodynamic simulations.

~\\
{\bf Sensitivity to nuclear structure}

Experimental measurements as well as event-by-event calculations have reached a degree of precision that has been demonstrated to be sensitive to the detailed nuclear structure of the colliding nuclei. Especially in ultra-central collisions, where even the elliptic deformation of the fireball is dominated by fluctuations, a strong sensitivity of $v_n$ coefficients to the nuclear deformation has been observed. For example, comparing ultra-central Au+Au collisions to those of the strongly deformed uranium nuclei \cite{Heinz:2004ir, Fortier:2024yxs}, the STAR Collaboration has extracted observables that show clear signals of the uranium deformation \cite{STAR:2015mki}.

Similar effects are seen at the LHC in Xe+Xe collisions \cite{Giacalone:2017dud, ALICE:2018lao}, and more recently collaborations between the members of the nuclear structure and heavy-ion communities have emerged \cite{Jia:2022ozr}. There is a strong synergy here as, in particular for {\it smaller} nuclei, ab-initio nuclear structure calculations can be used to reduce uncertainties in the initial state of the collision fireball, enabling tighter constraints on the QGP medium properties. Tighter medium constraints can then be exploited to analyze heavy-ion collision experiments with {\it larger} deformed nuclei, whose shape is less accurately known from nuclear structure theory, to obtain, via Bayesian inference, tighter constraints on the ground state deformation of these nuclei before the collision. Interesting targets of such a program include the neutron skin of Pb \cite{Nijs:2020ors, Giacalone:2023cet}, the hexadecapole deformation of $^{238}$U \cite{Ryssens:2023fkv}, or the triaxial structure of Xe isotopes around $^{129}$Xe \cite{Bally:2021qys, Zhao:2024lpc}.

~\\
{\bf Viscous relativistic hydrodynamic theory and numerical implementations}

The application of hydrodynamics to heavy-ion collisions and small systems has been a major driver for research efforts into the theoretical foundations of relativistic hydrodynamics. Seeking an answer to the question how hydrodynamics can be so successful in describing flow observables in systems of such extremely small sizes and short lifetimes has resulted in detailed analyses of various theories and their hydrodynamic limits. Ranging from kinetic theory to holography, these studies have revealed the existence of hydrodynamic attractors \cite{Heller:2015dha} and highlighted the relevance of whether hydrodynamic modes or non-hydrodynamic modes dominate the dynamics \cite{Florkowski:2017olj}. Comparisons of the attractors characterizing the evolution of the macroscopic hydrodynamic fields within microscopic descriptions (such as kinetic theory) with those of various different macroscopic hydrodynamic approximations have led to the identification of several far-off-equilibrium extensions of standard viscous relativistic fluid dynamics (e.g. viscous anisotropic hydrodynamics (aHydro) and Maximum Entropy hydrodynamics (ME hydro) that improve the accuracy of hydrodynamic simulations in specific non-equilibrium situations that are typical for high-energy heavy-ion collisions \cite{Heinz:2015gka, Jaiswal:2019cju, Jaiswal:2021uvv, Chattopadhyay:2023hpd}.

The development of numerical implementations of viscous relativistic hydrodynamics was strongly driven by the heavy-ion field. In many (albeit not all) astrophysical situations viscous effects are negligible, such that heavy-ion applications provided a stronger motivation for progress in this direction. Only recently, effects of bulk viscosity in neutron star mergers are being explored, see e.g.~\cite{Chabanov:2023blf}.

~\\
{\bf Thermalization}

Besides motivating research into better understanding the hydrodynamic limit of field theories, the success of hydrodynamic models triggered significant research into thermalization and isotropization within a wide range of theories. We laid out this progress in some detail in Section\,\ref{sec:therm}, discussing both weak and strong coupling approaches. One major insight has been that hydrodynamics can be applicable even when the system is not close to local thermal equilibrium, and the term ``hydrodynamization'' was coined for this situation. However, if one is not close to thermal equilibrium, the hydrodynamic description will in general deviate from the microscopic theory it is supposed to approximate -- the non-hydrodynamic modes present in both can differ \cite{Kurkela:2019set}.  Optimized far-off-equilibrium generalizations of viscous relativistic fluid dynamics, tailored to specific expansion geometries encountered in high-energy heavy-ion collisions, can address this issue with some success.

\section{Concluding remarks}
In closing, we emphasize that relativistic hydrodynamics has been essential to the scientific discoveries made in high energy heavy-ion collisions at RHIC and the LHC. Not only has its success allowed for the quantitative extraction of knowledge about the condensed matter properties of matter controlled by the strong interaction from the experimental measurements, but it has also stimulated and advanced exciting and important new theoretical developments. The field is a rich fountain of new ideas, with opportunities and synergies arising in collaborations with adjacent fields, including nuclear structure and astrophysics.

\begin{acknowledgement}
We thank Charles Gale, John Harris, Tetsufumi Hirano, Jaki Noronha-Hostler, and Urs Wiedemann for comments and useful discussions. This material is based upon work supported by the U.S. Department of Energy, Office of Science, Office of Nuclear Physics, under DOE Contract No.~\rm{DE-SC0012704} and Award No.~\rm{DE-SC0004286}.
\end{acknowledgement}
%

%%%%%%%%%%%%%%%%%%%%%%%% referenc.tex %%%%%%%%%%%%%%%%%%%%%%%%%%%%%%
% sample references
% %
% Use this file as a template for your own input.
%
%%%%%%%%%%%%%%%%%%%%%%%% Springer-Verlag %%%%%%%%%%%%%%%%%%%%%%%%%%
%
% BibTeX users please use
\bibliographystyle{spphys}
\bibliography{refs-inspire,refs}

%
%\biblstarthook{
%\section{Styling of References}

%\begin{thebibliography}{99.}%
% and use \bibitem to create references.
%
% Use the following syntax and markup for your references if 
% the subject of your book is from the field 
% "Mathematics, Physics, Statistics, Computer Science"
%
% Contribution 

% Journal article

% \bibitem{Cabibo_Parisi}  N. Cabibbo and G. Parisi: Exponential
%   Hadronic Spectrum and Quark Liberation, Phys.Lett. \textbf{B59}, 67--69 (1975).

%

%\end{thebibliography}

\end{document}